\documentclass {article}

\usepackage[T1] {fontenc} %formata melhor fontes acentuadas
\usepackage{graphicx}% Include figure files em eps, pdf e jpg
\usepackage[latin1]{inputenc}%tipo de acentos a serem reconhecidos: ã, á etc.

\begin {document}

\title {Hubble's Cosmology: From a Finite Expanding Universe to a Static Endless Universe\footnote {Paper accepted for presentation at the Second Crisis in Cosmology Conference, 2008 September 7-11, Port Angeles, Washington, USA.}}
\author {A. K. T. Assis$^1$\thanks {E-mail:
assis@if\/i.unicamp.br, homepage:
www.if\/i.unicamp.br/\~{}assis}, M. C. D. Neves$^2$\thanks {E-mail: macedane@yahoo.com, homepage: www.dfi.uem.br/\~{}macedane} \ and D. S. L. Soares$^3$\thanks {E-mail: dsoares@fisica.ufmg.br, homepage: www.fisica.ufmg.br/\~{}dsoares}\\
 \\
1 - Institute of Physics `Gleb Wataghin'\\
University of Campinas --- UNICAMP\\
13083-970 Campinas, SP, Brazil\\
 \\
2 - Departamento de F\'{\i}sica\\
Funda\c c\~ao Universidade Estadual de Maring\'a - FUEM\\
87020-900 Maring\'a, PR, Brazil\\
 \\
3 - Departamento de F\'{\i}sica, ICEx\\
Universidade Federal de Minas Gerais\\
C. P. 702, 30123-970 Belo Horizonte, MG, Brazil}
\date {}

\maketitle

{\bf Short Abstract ($<$ 60 words):} We analyze Hubble's approach to cosmology. In 1929 he accepted a finite expanding universe in order to explain the redshifts of distant galaxies. Later on he turned to an infinite stationary universe due to observational constraints. We show, by quoting his works, that he remained cautiously against the big bang until the end of his life. 

\vskip.5cm
{\bf Full Abstract (200-500 words):} We analyze the views of Edwin Hubble  
(1889-1953) as regards the large scale structure of the universe. In 1929 
he initially accepted a finite expanding universe in order to explain the 
redshifts of distant galaxies. Later on he turned to an infinite stationary 
universe and a new principle of nature in order to explain the same 
phenomena. Initially, he was impressed by the agreement of his 
redshift-distance relation with one of the predictions of de Sitter's 
cosmological model, namely, the so-called ``de Sitter effect'', the 
phenomenon of the scattering of material particles, leading to an 
expanding universe. A number of observational evidences, though, 
made him highly skeptical with such a scenario. They were better 
accounted for by an infinite static universe. The evidences he found 
were: (i) the huge values he was getting for the ``recession'' velocities 
of the nebulae (1,800 km/s in 1929 up to 42,000 km/s in 1942, leading 
to $v/c = 1/7$), with the redshifts interpreted as velocity-shifts. All 
other known real velocities of large astronomical bodies are much 
smaller than these. (ii) The ``number effect'' test, which is the running 
of nebulae luminosity with redshift. Hubble found that a static universe 
is, within the observational uncertainties, slightly favored. The test is 
equivalent to the modern ``Tolman effect'', for galaxy surface 
brightnesses, whose results are still a matter of dispute. (iii) The 
smallness of the size and the age of the curved expanding universe, 
implied by the expansion rate that he had determined, and, (iv) the 
fact that an uniform distribution of galaxies on large scales is more 
easily obtained from galaxy counts, when a static and flat  model 
is considered.  In an expanding and closed universe, Hubble found 
that homogeneity was only obtained at the cost of a large curvature. 
We show, by quoting his works, that Hubble remained cautiously 
against the big bang until the end of his life, contrary to the 
statements of many modern authors. In order to account for 
redshifts, in a non-expanding universe, Hubble called for a new 
principle of nature, like the ``tired-light'' mechanism proposed by 
Fritz Zwicky in 1929. On the other hand, he was aware of the 
theoretical difficulties of such a radical assumption. Hubble's 
approach to cosmology strongly suggests that he would not agree 
with the present status of the modern cosmological paradigm, 
since he was, above all, driven by observations and by the 
consequences derived from them.

\vfill


\vfill


\vfill\eject

\section {Introduction}

The concept of an expanding universe, as we are familiar with nowadays, was invented independently by the Russian scientist Alexander Friedmann and by the Belgian cosmologist Georges Lema\^\i tre, with their solutions of Einstein's theory of general relativity applied to the cosmic fluid.  Their pioneering papers on the subject were published in 1922 and 1924 (AF), as well as 1927 and 1931 (GL). The redshift-apparent magnitude (or distance) relation discovered by Edwin Hubble in 1929, \cite {hubble29}, fitted nicely the new theoretical picture. The so-called {\it Hubble's law} was precisely the one predicted by both Friedmann and Lema\^\i tre's models. It was immediately raised to the status of an {\it ``observational'' discovery of the expanding universe}. 

This, of course, is not the case. The idea of an expansion is, first of all, a theoretical idea --- a strange ``effect'' in the early empty de Sitter's model. Hubble's observations are consistent with the idea, but are not necessarily a proof of it. Hubble himself was aware of this and sought during all his life for the correct answer to the question posed by his discovery: {\it What does cause redshifts?} The two possibilities considered by him were 
the expanding relativistic models and the tired-light paradigm. The latter, incidentally, is not characterized by a particular physical theory, which has not yet been found. The idea was originally suggested by one of Hubble's greatest friends, Fritz Zwicky, \cite{zwicky29}. He and Richard Tolman are gratefully acknowledged in the preface of Hubble's {\it The Realm of the Nebulae} with these impressive words, \cite [p. ix] {hubble58}: {\it In the field of cosmology, the writer has had the privilege of consulting Richard Tolman and Fritz Zwicky of the California Institute of Technology. Daily contact with these men has engendered a common atmosphere in which ideas develop that cannot always be assigned to particular sources. The individual, in a sense, speaks for the group}. By the way, this is the aim of any successful collaboration.  
 
In his {\it Principia}, Book III, Isaac Newton presented the ``Rules of Reasoning in Philosophy''. His third rule reads as follows, \cite [p. 398] {newton34}: {\it The qualities of bodies, which admit neither intensification nor remission of degrees, and which are found to belong to all bodies within the reach of our experiments, are to be esteemed the universal qualities of all bodies whatsoever.} In his comments on this rule he stated that 
``We are certainly not to relinquish the evidence of experiments for the sake of dreams and vain fictions of our own devising''. As we show below, this requirement was definitely fulfilled by Hubble's approach to cosmology.

\section {Hubble and His Changing Views about Cosmology}

Edwin Powell Hubble (1889-1953) established in 1924 that many nebulae are stellar systems outside the Milky Way, when he discovered Cepheid variables in the Andromeda Nebula using the 100-inch telescope on Mount Wilson.

In 1929 he established the famous distance-velocity relation which is also called nowadays the {\it law of redshift} or {\it Hubble's law}, \cite {hubble29}. The title of his paper reads: ``A relation between distance and radial velocity among extra-galactic nebulae.'' In Table I Hubble presented a symbol {\it v} and called it the ``measured velocities in km./sec.'' As a matter of fact he and his collaborator Milton L. Humason (1891-1972) never {\it measured} velocities directly. What they measured were the redshifts of these extra-galactic nebulae. But in this crucial paper Hubble considered these redshifts as representing real radial velocities of these nebulae. The main conclusion of the paper appeared on page 139: ``The data in the table indicate a linear correlation between distances and velocities, whether the latter are used directly or corrected for solar motion, according to the older solutions.'' The latter paragraph of the paper presented Hubble's interpretation of his findings, the meaning he gave them in 1929, namely:

\begin {quote}
The outstanding feature, however, is the possibility that the velocity-distance relation may represent the de Sitter effect, and hence that numerical data may be introduced into discussions of the general curvature of space. In the de Sitter cosmology, displacements of the spectra arise from two sources, an apparent slowing down of atomic vibrations and a general tendency of material particles to scatter. The latter involves an acceleration and hence introduces the element of time. The relative importance of these two effects should determine the form of the relation between distances and observed velocities; and in this connection it may be emphasized that the linear relation found in the present discussion is a first approximation representing a restricted range in distance.
\end {quote}

Willem de Sitter (1872-1934) was a Dutch mathematician, physicist and astronomer. In 1916-17 he had found a solution to Einstein's field equations of general relativity describing the expansion of the universe. Hubble had met him in 1928 in Leiden, where de Sitter was both professor of astronomy at the University of Leiden and director of the Leiden Observatory, \cite [p. 198] {christianson96}. This last paragraph of Hubble's paper shows that in 1929 he thought of the expansion of the universe as a real possibility.

However, it must be pointed out that, as early as 1935, Hubble was much more cautious when referring to velocities of recession. In a paper with R. Tolman, \cite{hubbletolman35}, already in the introductory section, the authors made a plain statement of their worries about the proper nomenclature:

\begin{quote}
Until further evidence is available, both the present writers wish to express an open mind with respect to the ultimately most satisfactory explanation of the nebular red-shift and, in the presentation of purely observational findings, to continue to use the phrase ``apparent'' velocity of recession. They both incline to the opinion, however, that if the red-shift is not due to recessional motion, its explanation will probably involve some quite new physical principles. 
\end{quote}

Hubble was invited to deliver eight Silliman Lectures at Yale University in 1935. These lectures formed the basis of his book {\it The Realm of the Nebulae}, published in 1936. In this book he was more careful in stating what was really measured and what was interpreted. On pages 2 and 3 he mentioned that the observer accumulates data of apparent luminosities of nebulae and red-shifts of their spectra. The distances of these nebulae may be indicated by their faintness, as these distances are not measured directly. The simplest relation obtained between these two data is ``a linear relation between red-shifts and distances as indicated by the faintness of the nebulae,'' \cite [p. 3] {hubble58}.

As regards the origins of these red-shifts, that is, what causes them, he mentioned on pages 33-34 that a possible {\it interpretation} is that they might be due to radial motion of these nebulae with respect to the Earth: 

\begin {quote}
Nebular spectra are peculiar in that the lines are not in the usual positions found in nearby light sources. They are displaced toward the red end of their normal position, as indicated by suitable comparison spectra. The displacements, called red-shifts, increase, on the average, with the apparent faintness of the nebula that is observed. Since apparent faintness measures distance, it follows that red-shifts increase with distance. Detailed investigation shows that the relation is linear.

Small microscopic shifts, either to the red or to the violet, have long been known in the spectra of astronomical bodies other than nebulae. These displacements are confidently interpreted as the results of motion in the line of sight --- radial velocities of recession (red-shifts) or of approach (violet-shifts). The same interpretation is frequently applied to the red-shifts in nebular spectra and has led to the term ``velocity-distance'' relation for the observed relation between red-shifts and apparent faintness. On this assumption, the nebulae are supposed to be rushing away from our region of space, with velocities that increase directly with distance.

Although no other plausible explanation of red-shifts has been found, the interpretation as velocity-shifts may be considered as a theory still to be tested by actual observations. Critical tests can probably be made with existing instruments. Rapidly receding light sources should appear fainter than stationary sources at the same distances, and near the limits of telescopes the ``apparent'' velocities are so great that the effects should be appreciable.
\end {quote}

He was utilizing the adjective ``apparent'' before the word ``velocities'' in order to emphasize that this was only an interpretation. Hubble himself began to propose possible tests in order to verify or to falsify this assumption, as indicated by the last paragraph above. Chapter V of this book is devoted to ``the velocity-distance relation.'' In Plate VIII of the book Hubble said: ``Red-shifts resemble velocity-shifts, and no other satisfactory explanation is available at the present time: red-shifts are due either to actual motion of recession or to some hitherto unrecognized principle of physics.'' On pages 121-3 he presented cautious remarks about the interpretation of these red-shifts, our emphasis:

\begin {quote}
Observations show that details in nebular spectra are displaced toward the red from their normal positions, and that the red-shifts increase with apparent faintness of the nebulae. Apparent faintness is confidently interpreted in terms of distance. Therefore, the observational result can be restated --- red-shifts increase with distance.

{\it Interpretations of the red-shifts themselves do not inspire such complete confidence}. Red-shifts may be expressed as fractions, $d\lambda /\lambda$, where $d\lambda$ is the displacement of a spectral line whose normal wave-length is $\lambda$. The displacements, $d\lambda$, vary systematically through any particular spectrum, but the variation is such that the fraction, $d\lambda /\lambda$, remains constant. Thus $d\lambda /\lambda$ specifies the shift for any nebula, and it is the fraction which increases linearly with distances of the nebulae.\footnote {[Note by Hubble:] The apparent radial velocity of a nebula is, to a first approximation, the velocity of light (186,000 miles/sec.) multiplied by the fraction, $d\lambda /\lambda$.} From this point, the term red-shift will be employed for the fraction $d\lambda /\lambda$.

Moreover, the displacements, $d\lambda$, are always positive (toward the red) and so the wave-length of a displaced line, $\lambda + d\lambda$, is always greater than the normal wavelength, $\lambda$. Wave-lengths are increased by the factor $(\lambda + d\lambda)/\lambda$, or the equivalent $1 + d\lambda/\lambda$. Now there is a fundamental relation in physics which states that the energy of any light quantum, multiplied by the wavelength of the quantum, is constant. Thus

\centerline {Energy $\times$ wave-length = constant.}

Obviously, since the product remains constant, red-shifts, by increasing wave-lengths, must reduce the energy in the quanta. Any plausible interpretation of red-shifts must account for the loss of energy. The loss must occur either in the nebulae themselves or in the immensely long paths over which the light travels on its journey to the observer.

Thorough investigation of the problem has led to the following conclusions. Several ways are known in which red-shifts might be produced. Of them all, only one will produce large shifts without introducing other effects which should be conspicuous, but which are not observed. This explanation interprets red-shifts as Doppler effects, that is to say, as velocity-shifts, indicating actual motion of recession. {\it It may be stated with some confidence that red-shifts are velocity-shifts or else they represent some hitherto unrecognized principle in physics.}

The interpretation as velocity-shifts is generally adopted by theoretical investigators, and the velocity-distance relation is considered as the observational basis for theories of an expanding universe. Such theories are widely current. They represent solutions of the cosmological equation, which follow from the assumption of a nonstatic universe. They supersede the earlier solutions made upon the assumption of a static universe, which are now regarded as special cases in the general theory.

Nebular red-shifts, however, are on a very large scale, quite new in our experience, and empirical confirmation of their {\it provisional interpretation as familiar velocity-shifts}, is highly desirable. Critical tests are possible, at least in principle, since rapidly receding nebulae should appear fainter than stationary nebulae at the same distances. The effects of recession are inconspicuous until the velocities reach appreciable fractions of the velocity of light. This condition is fulfilled, and hence the effects should be measurable, near the limits of the 100-inch reflector.

The problem will be discussed more fully in the concluding chapter. The necessary investigation are beset with difficulties and uncertainties, and conclusions from data now available are rather dubious. They are mentioned here in order to emphasize the fact that the interpretation of red-shifts is at least partially within the range of empirical investigation. For this reason the attitude of the observer is somewhat different from that of the theoretical investigator. {\it Because the telescopic resources are not yet exhausted, judgment may be suspended until it is known from observations whether or not red-shifts do actually represent motion.}

{\it Meanwhile, red-shifts may be expressed on a scale of velocities as a matter of convenience. They behave as velocity-shifts behave and they are very simply represented on the same familiar scale, regardless of the ultimate interpretation. The term ``apparent velocity'' may be used in carefully considered statements, and the adjective always implied where it is omitted in general usage.}
\end {quote}

The test proposed by Hubble was called ``the number-effect.'' He described it on pages 193-196 of {\it The Realm of the Nebulae}:

\begin {quote}
The effects of red-shifts are calculated on the alternative assumptions that (a) they represent motion (are velocity-shifts) and (b) they do not represent motion. Since the numerical results are not the same, the observed departures may be used to identify the correct interpretation. [...]

Radiation from a nebula may be pictured as light-quanta-parcels of energy --- streaming out in all directions. Apparent luminosity is measured by the rate at which the quanta reach the observer, together with the energy in the quanta. If either the energy or the rate of arrival is reduced, the apparent luminosity is diminished. Red-shifts reduce the energy in the quanta whether the nebulae are stationary or receding. Thus an ``energy-effect'' may be expected, regardless of the interpretation of red-shifts. {\it The rate of arrival (i.e., the number of quanta reaching the observer per second) is reduced if the nebulae are receding from the observer, but not otherwise. This phenomenon, known as the ``number-effect,'' should in principle furnish a crucial test of the interpretation of red-shifts as velocity-shifts.} 
\end {quote}

The ``number-effect'', more precisely, the ``number-of-photons effect'', was in fact treated in an earlier paper with his friend and long-time collaborator, the cosmologist Richard Tolman, \cite {hubbletolman35} and references therein, who originally proposed such a test. The test is the so-called {\it Tolman effect}, which was thoroughly investigated later on by Sandage and co-workers (e.g. \cite {lubinsandage01} and references therein). A positive result for the reality of the expansion by means of such a test is still not definitive because observational uncertainties and evolutionary effects jeopardize the final conclusion (but see Andrews 2006 \cite {andrews06} and Lerner 2006
\cite {lerner06}, who found a negative result for the test). 

Hubble's preliminary conclusion coming from observations was clearly against the interpretation that the red-shifts are due to the radial motion of the nebulae away from the Earth, \cite [p. 197] {hubble58}:

\begin {quote}
The observed coefficient [of magnitude-increment] is smaller here than that in the relation calculated on either interpretation of red-shifts, but is much closer to the coefficient representing no motion. Careful examination of possible sources of uncertainties suggests that the observations can probably be accounted for if red-shifts are not velocity-shifts. If redshifts are velocity-shifts then some vital factors must have been neglected in the investigation.
\end {quote}

In the same year in which {\it The Realm of the Nebulae} was published, 1936, Hubble delivered three Rhodes Memorial Lectures in Oxford on October 29, November 12 and 26. From a note published in Nature we can know the points of view expressed by Hubble in these lectures, our emphasis, \cite {hhp36}:

\begin {quote}
[...] The lectures, which dealt in turn with the observable region, the role of the red-shifts, and possible models of the universe, have revealed that a static universe with a hitherto unsuspected dependence of light frequency on distance is probably more acceptable than one or other of the homogeneous expanding models of general relativity.

[...] Without in any way straining the observations, but at the expense of a newly postulated property of radiation, we can describe the nebular counts in terms of a simple static universe. [...]

If no recession is assumed, the observed nebular counts are satisfactorily described by supposing that we are observing a finite portion of a much larger universe of nebulae, but a universe in which the frequency of light varies uniformly with the distance. If, on the other hand, recession is assumed, the observed nebular counts are not satisfactorily described by any of the homogeneous expanding models of general relativity, but if forced to fit require that the universe be closed, that we have already explored it to its outmost bounds with the 100-in. telescope, and that it is a universe dominantly filled with non-luminous matter distributed in such a way as to absorb or scatter negligibly small amounts of light.

{\it The large and appreciative audiences who followed the three lectures, each a model of exposition and clarity, had little difficulty in agreeing with Dr. Hubble that the consequences of assuming no recession were the less difficult to accept.}
\end {quote}

These lectures were published in 1937 under the title {\it The Observational Approach to Cosmology}, \cite {hubble37}. In this book Hubble mentioned that what led him to look for alternative interpretations for the red-shifts of the nebulae, instead of the usual interpretation as being due to an actual radial velocity away from the Earth, was the large values these apparent velocities were reaching, \cite [p. 29] {hubble37}: ``The disturbing features were the facts that the `velocities' reached enormous values and were precisely correlated with distance.'' 

It is worthwhile to comment a little on this important aspect mentioned by Hubble. In 1929 the largest value of the radial velocity $v$ of the nebulae quoted by Hubble was $v = 1,800$ km/s, \cite [Table 2] {hubble29}. This implies
$v/c = 6\times 10^{-3}$, where $c = 3\times 10^8$ m/s is the light velocity in vacuum. By 1936 when he wrote {\it The Realm of the Nebulae} this value had increased to $v = 39,000$ km/s, implying $v/c = 0.13$, \cite [Plate VII, p. 104] {hubble58}. In 1942 he was obtaining radial velocities of recession to 1/7 the velocity of light, implying $v/c \approx 0.14$, \cite [p. 104] {hubble42}. These extremely large recession velocities are a source of doubt for the interpretation of the redshift as a velocity effect. The reason is that all other velocities of large astronomical objects known to us are much smaller. For instance, the orbital velocity of the Earth around the Sun is approximately 30 kms$^{-1}$ ($v/c \approx 10^{-4}$); the orbital velocity of the solar system relative to the center of our galaxy is approximately 250 kms$^{-1}$ ($v/c \approx 10^{-3}$); and the random or peculiar motion of galaxies is of this same order of magnitude.

In order to remove the disturbing features of extremely high `velocities' of recession, Hubble presented on page 30 of his book {\it The Observational Approach to Cosmology} a more plausible interpretation, our emphasis: 

\begin {quote}
Well, perhaps the nebulae are all receding in this peculiar manner. But the notion is rather startling. The cautious observer naturally examines other possibilities before accepting the proposition even as a working hypothesis. He recalls the alternative formulation of the law of red-shifts --- light loses energy in proportion to the distance it travels through space. {\it The law, in this form, sounds quite plausible. Internebular space, we believe, cannot be entirely empty.} There must be a gravitational field through which the light-quanta travel for many millions of years before they reach the observer, and {\it there may be some interaction between the quanta and the surrounding medium}. The problem invites speculation, and, indeed, has been carefully examined. But no satisfactory, detailed solution has been found. The known reactions have been examined, one after the other --- and they have failed to account for the observations. Light {\it may} lose energy during its journey through space, but if so, we do not yet know how the loss can be explained.
\end {quote}

On the third Chapter of this book, page 45, he summarized possible alternative explanations as follows, our emphasis: ``The previous lecture described the appearance and behaviour of red-shifts in the spectra of nebulae, and called attention to the alternative possible interpretations. If red-shifts are produced in the nebulae, where the light originates, they are probably the familiar velocity-shifts, and they measure an expansion of the universe. If the nebulae are not rapidly receding, red-shifts are probably introduced between the nebulae and the observer; {\it they represent some unknown reaction between the light and the medium through which it travels.}'' In the next page he expressed clearly his suspicions against the expansion of the universe, namely: ``The assumption of motion, on the other hand, led to a non-linear law of red-shifts, according to which the velocities of recession accelerate with distance or with time counted backward into the past. A universe that has been expanding in this manner would be so extraordinarily young, the time-interval since the expansion began would be so brief, that suspicions are at once aroused concerning either the interpretation of red-shifts as velocity-shifts or the cosmological theory in its present form.''

When the law of nebular distribution was not interpreted as velocity-shifts, Hubble obtained a uniform distribution of nebulae and was very satisfied, \cite [p. 49] {hubble37}: ``The uniform distribution is a plausible and welcome result.'' On page 51 he went on: ``Therefore, we accept the uniform distribution, and assume that space is sensibly transparent. Then the data from the surveys are simply and fully accounted for by the energy corrections alone --- without the additional postulate of an expanding universe.'' On pages 60-61 he presented another dubious conclusion arising from the assumption of an expanding universe, our emphasis:

\begin {quote}
The nature of the [spatial] curvature has rather grave implications. Since the curvature is positive, the universe is closed. Space is closed as the surface of a sphere is closed. The universe has a definite, finite volume although it has no boundaries in three-dimensional space. The remarkably small numerical value of the radius of curvature is a complete surprise. It implies that a large fraction of the universe, perhaps a quarter, can be explored with existing telescopes.\footnote {[Note by Hubble:] The volume of this universe would be $2\pi^2R^3$, where $R$ is the radius of curvature, or about $2\times 10^{27}$ cubic light-years. The universe would contain about 400 million nebulae.} {\it The small volume of the universe is another strange and dubious conclusion}. The familiar interpretation of red-shifts as velocity-shifts very seriously restricts not only the time-scale, the age of the universe, but the spatial dimensions as well. On the other hand, the alternative possible interpretation, that red-shifts are not velocity-shifts, avoids both difficulties, and presents the observable region as an insignificant sample of a universe that extends indefinitely in space and in time.
\end {quote}

At the end of the book he presented clearly his preferred model of the universe, our emphasis, \cite [pp. 63-64] {hubble37}:

\begin {quote}
[...] {\it Nevertheless, the ever-expanding model of the first kind seems rather dubious. It cannot be ruled out by observations, but it suggests a forced interpretation of the data.}

The disturbing features are all introduced by the recession factors, by the assumption that red-shifts are velocity-shifts. The departure from a linear law of red-shifts, the departure from uniform distribution, the curvature necessary to restore homogeneity, the excess material demanded by the curvature, each of these is merely the recession factor in another form. These elements identify a unique model among the array of possible expanding worlds, and, in this model, the restriction in the time-scale, the limitation of the spatial dimensions, the amount of unobserved material, is each equivalent to the recession factor.

{\it On the other hand, if the recession factor is dropped, if red-shifts are not primarily velocity-shifts, the picture is simple and plausible.} There is no evidence of expansion and no restriction of the time-scale, no trace of spatial curvature, and no limitation of spatial dimensions. Moreover, there is no problem of internebular material. The observable region is thoroughly homogeneous; it is too small a sample to indicate the nature of the universe at large. The universe might even be an expanding model, provided the rate of expansion, which pure theory does not specify, is inappreciable. For that matter, the universe might even be contracting.
\end {quote}

It is very easy to know which one of the two pictures of the universe was that one preferred by Hubble himself. This is the choice he presented in the last paragraph of this book, \cite [p. 66] {hubble37}: ``Two pictures of the universe are sharply drawn. Observations, at the moment, seem to favour one picture, but they do not rule out the other. We seem to face, as once before in the days of Copernicus, a choice between a small, finite universe, and a universe indefinitely large plus a new principle of nature.''

Interesting discussions of this endless universe without expansion have been given by Reber and Marmet, \cite {reber77}, \cite {reber86} and \cite {marmetreber89}.

Five years later Hubble returned to this subject presenting essentially the same points of view, although with more data, in a paper called ``The problem of the expanding universe,'' \cite {hubble42}. He expressed his goal as follows: ``One phase of this ambitious project is the observational test of the current theory of the expanding universes of general relativity.'' He presents the usual interpretation of the red-shifts as velocity-shifts. He mentions that ``the observations have been carried out to nearly 250 million light years where the red shifts correspond to velocities of recession of nearly 25,000 miles per second or 1/7 the velocity of light.'' As we saw above, Hubble was disturbed by these enormous values. After presenting the consequences of this usual interpretation, he commented as follows: ``This pattern of history seems so remarkable that some observers view it with pardonable reserve, and try to imagine alternative explanations for the law of red shifts. Up to the present, they have failed. Other ways are known by which red shifts might be produced, but all of them introduce additional effects that should be conspicuous and actually are not found. Red shifts represent Doppler effects, physical recession of the nebulae, or the action of some hitherto unrecognized principle in nature.'' He compared the theory of expansion with the actual observations of nebulae and concluded as follows: ``The remainder of the recently accumulated information is not favorable to the theory. It is so damaging, in fact, that the theory, in its present form, can be saved only by assuming that the observational results include hidden systematic errors.'' In a section devoted to the interpretation of red shifts he mentioned that his ``investigations were designed to determine whether or not red shifts represent actual recession.'' In the first and second Figures of this paper he showed how a stationary universe gave a better fit to the data than an expanding universe as regards the law of red shifts and the large scale distribution of nebulae. As regards this last aspect, he concluded as follows: ``On the assumption that red shifts do not represent actual recession, the large scale distribution is sensibly homogeneous --- the average number of nebulae per unit volume of space is much the same for each of the spheres. [...] All of these data lead to the very simple conception of a sensibly infinite, homogeneous universe of which the observable region is an insignificant sample.'' 

The first and last paragraphs of his conclusion are very clear as regards his preferred model of the universe and should be quoted in full, namely:

\begin {quote}
Thus the use of dimming corrections leads to a particular kind of universe, but one which most students are likely to reject as highly improbable. Furthermore, the strange features of this universe are merely the dimming corrections expressed in different terms. Omit the dimming factors, and the oddities vanish. We are left with the simple, even familiar concept of a sensibly infinite universe. All the difficulties are transferred to the interpretation of red shifts which cannot then be the familiar velocity shifts.

[...]

Meanwhile, on the basis of the evidence now available, apparent discrepancies between theory and observations must be recognized. A choice is presented, as once before in the days of Copernicus, between a strangely small, finite universe and a sensibly infinite universe plus a new principle of nature.
\end {quote}

A possibility of what he thought this new principle of nature might be can be found in an interview published in 1948, when he was the cover of the Time magazine, \cite {time48}: ``Other critics question the ``red shift'' as a measure of velocity. The usual explanation of the reddening effect is that the luminous body's motion away from the observer ``pulls out'' the light waves, making them longer (redder) than normal. But since red light contains less energy per unit (photon) than violet light, Hubble's critics suggest that light may lose some of its energy in traversing space, thus turning redder. It may start out from a distant nebula as young, vigorous violet and arrive at the earth after millions of weary years as old, tired red. If that is what happens, perhaps nebulae are not moving at all? [...] Meanwhile, he [Hubble] will look for evidence that the ``red shift'' does not indicate speed but is due to some other effect, such as light getting ``tired.'' Hubble does not expect such evidence, but will welcome it if he finds it. Tired light, he thinks, would be a discovery quite as sensational as the exploding universe.''

Christianson quoted this interview and expressed clearly how Hubble thought about this possibility, \cite [p. 318] {christianson96}:

\begin {quote}
Hubble then addressed an alternative hypothesis embraced by those who found the expansion theory too fantastic. The redshift, it was argued, does not indicate expansion but something quite different. Light starts out from a distant nebula as young, vigorous, and violet. But after millions of years its energy is depleted, its waves elongate, and it turns redder, transforming it into the ``tired light'' captured on the plates taken at Mount Wilson and Palomar. If this is what happens, the nebulae may be moving very little --- or not at all.

While Hubble would not be pushed into a corner, he finally admitted that he did ``not expect'' to find visual evidence that would undermine the redshift hypothesis, yet he would ``welcome it if he finds it. Tired light... would be a discovery quite as sensational as the exploding universe.''
\end {quote}

\section {Concluding remarks}

Hubble was strongly driven by observations in his conclusions against the expanding closed universe. Especially related to this particular issue was the value of the expansion rate, i.e., {\it Hubble's constant}, which was  determined by himself in 1929 as $\approx$ 500 kms$^{-1}$Mpc$^{-1}$, \cite {hubble29}. Decades later, after Hubble's death in 1953, the value was revised to the well-known range 50-100 kms$^{-1}$Mpc$^{-1}$ (or more specifically 72 kms$^{-1}$Mpc$^{-1}$, according to the final results of the HST Key Project on H$_\circ$, \cite{freedmanetal01}).

With an almost tenfold smaller recession factor and, consequently, an almost tenfold larger age for the universe, the model would seem much more palatable as regards its space and time dimensions. 

But at what cost? In order to save the big bang model, it is necessary to introduce innumerable ad hoc hypotheses. These many hypotheses reminds us of a story told by Lakatos, \cite [pp. 100-1] {lakatos70}:

\begin {quote}
The story is about an imaginary case of planetary misbehavior. A physicist of the pre-Einsteinian era takes Newton's mechanics and his law of gravitation, $N$, the accepted initial conditions, $I$, and calculates, with their help, the path of a newly discovered small planet, $p$. But the planet deviates from the calculated path. Does our Newtonian physicist consider that the deviation was forbidden by Newton's theory and therefore that, once established, it refutes the theory $N$? No. He suggests that there must be a hitherto unknown planet $p'$ which perturbs the path of $p$. He calculates the mass, orbit, etc., of this hypothetical planet and then asks an experimental astronomer to test his hypothesis. The planet $p'$ is so small that even the biggest available telescopes cannot possible observe it: the experimental astronomer applies for a research grant to build yet a bigger one. In three years' time the new telescope is ready. Were the unknown planet $p'$ to be discovered, it would be hailed as a new victory of Newtonian science. But it is not. Does our scientist abandon Newton's theory and his idea of the perturbing planet? No. He suggests that a cloud of cosmic dust hides the planet from us. He calculates the location and properties of this cloud and asks for a research grant to send up a satellite to test his calculations. Were the satellite's instruments (possibly new ones, based on a little-tested theory) to record the existence of the conjectural cloud, the result would be hailed as an outstanding victory for Newtonian science. But the cloud is not found. Does our scientist abandon Newton's theory, together with the idea of the perturbing planet and the idea of the cloud which hides it? No. He suggests that there is some magnetic field in that region of the universe which disturbed the instruments of the satellite. A new satellite is sent up. Were the magnetic field to be found, Newtonians would celebrate a sensational victory. But it is not. Is this regarded as a refutation of Newtonian science? No. Either yet another ingenious auxiliary hypothesis is proposed or... the whole story is buried in the dusty volumes of periodicals and the story never mentioned again.
\end {quote}

The innumerable ad hoc hypotheses introduced in the big bang model, in order to make it consistent with the matter-energy content of the real universe, only suggests what a real-world driven Edwin Hubble might have thought of it. 

%\section* {Acknowledgments:} ...

\newpage

%\bibliography {bibl}
%\bibliographystyle {unsrt}

\end {document}